\documentclass[10pt,letterpaper]{article}

\usepackage{amsmath,amssymb}

\usepackage{changepage}

\usepackage[utf8x]{inputenc}

\usepackage{textcomp,marvosym}

\usepackage{cite}

\usepackage{nameref,hyperref}

\usepackage{array}

\usepackage{setspace} 
\usepackage[aboveskip=1pt,labelfont=bf,labelsep=period,justification=raggedright,singlelinecheck=off]{caption}

\usepackage{graphicx}

\usepackage{algorithm}
\usepackage[noend]{algpseudocode}

\usepackage{braket}

\usepackage[draft]{changes}
\definechangesauthor[color=red]{rs}
\definechangesauthor[color=blue]{ya}
\definechangesauthor[color=green]{is}

\begin{document}
\vspace*{0.2in}

\begin{flushleft}
{\Large
\textbf\newline{Network community detection on small quantum computers} 
}
\newline
\\
Ruslan Shaydulin\textsuperscript{1,*},
Hayato~Ushijima-Mwesigwa\textsuperscript{1},
Ilya~Safro\textsuperscript{1},
Susan~Mniszewski\textsuperscript{2},
Yuri~Alexeev\textsuperscript{3}
\\
\bigskip
\textbf{1} School of Computing, Clemson University, Clemson, SC, USA 29634
\\
\textbf{2} Computer, Computational, \& Statistical Sciences Division, Los Alamos National Laboratory, Los Alamos, NM, USA 87545
\\
\textbf{3} Computational Science Division, Argonne National Laboratory, Argonne, IL, USA 60439
\\
\bigskip

* rshaydu@g.clemson.edu

\end{flushleft}
\section*{Abstract}
In recent years a number of quantum computing devices with small numbers of qubits became available. We present a hybrid quantum local search (QLS) approach that combines a classical machine and a small quantum device to solve problems of practical size. The proposed approach is applied to the network community detection problem. QLS is hardware-agnostic and easily extendable to new quantum computing devices as they become available. We demonstrate it to solve the 2-community detection problem on graphs of size up to 410 vertices using the 16-qubit IBM quantum computer and D-Wave 2000Q, and compare their performance with the optimal solutions.  Our results demonstrate that QLS perform similarly in terms of quality of the solution and the number of iterations to convergence on both types of quantum computers and it is capable of achieving results comparable to state-of-the-art solvers in terms of quality of the solution including reaching the optimal solutions. \\
\noindent {\bf Reproducibility}: Our code and data are available at \url{http://bit.ly/QLSCommunity}.

\section*{Introduction}
The recent years saw rapid progress in development of quantum computing (QC) devices. Multiple paradigms have been proposed and implemented in hardware introducing a variety of limitations that must be addressed prior to the wide application of QC. In particular, noisy intermediate scale quantum (NISQ) devices are widely expected to be limited up to a few hundred, or perhaps a few thousand qubits~\cite{preskill2018quantum}, severely restricting the size of the problems that can be tackled directly. As the potential of these NISQ-era quantum devices is becoming evident~\cite{dunjko2018computational}, there is an increasing interest in developing algorithms that leverage the small quantum devices that are becoming available. This requires the use of hybrid quantum-classical approaches where a problem is solved across
a CPU and a QC device.

The number of qubits in NISQ-era devices available at the time of writing is not nearly enough to demonstrate quantum advantage, which makes it especially hard to demonstrate the usefulness of quantum computers to solve real problems. For example, the possibility of quantum speedup using the hybrid quantum approximate optimization algorithm (QAOA) for a network 
problem similar to the one discussed in this paper (max-cut) is a subject of active discussion. On one hand, there are theoretical results demonstrating that QAOA for max-cut problem improves upon best know classical approximation algorithms for certain graphs~\cite{osti_1492737, PhysRevA.97.022304}. At the same time, there are indications that achieving speedup using QAOA might require at least several hundred qubits~\cite{guerreschi2018qaoa}. Research and development of quantum algorithms is necessary as the number and quality of qubits is improving.
These quantum algorithms can also be used to improve classical algorithms~\cite{tang2018quantum}. The need for development of new quantum algorithms was highlighted in the recent National Academy of Science report~\cite{nas2018quantum}. One of the important directions to make quantum computing feasible in the near future is to use various problem decomposition approaches to solve a large problems as a set of subproblems. This can be accomplished at various levels such as problem formulation or at the algorithmic level as demonstrated in this paper.

The decomposition approach might be the key method to achieve a quantum speedup  on even modest-size NISQ devices in near-term future. To support this claim, there is an important and encouraging work~\cite{bian2016mapping}, where it was shown that large combinatorial optimization problems can be effectively decomposed into subproblems on quantum annealing hardware, while still obtaining high quality of the overall solution. It was demonstrated for solving embedding problems on D-Wave quantum computers, but we believe that the same technique can be used to improve dramatically the speed and performance of QAOA algorithms on universal quantum computers.

In this work, we introduce the quantum local search (QLS) algorithm for the network community detection problem that is based on the local search method~\cite{Aarts:1997:LSC:549160}. Many different versions of the local search have been applied to numerous  computationally hard problems such as the  satisfiability testing \cite{selman1993local}, and the traveling salesman problem \cite{wu2015local,johnson1990local}. Local search is used for problems where a global solution cannot be computed directly but instead can be iteratively approximated in the space of candidate solutions (sub-problems), until optimal (or sufficiently good).
The important feature of QLS is that it is a hybrid hardware-agnostic algorithm that combines a classical machine with a small quantum device. In this method, QLS allows us to leverage available NISQ-era quantum devices to solve machine learning problems of practical size for the first time. 

A version of the network community detection (also known as graph clustering)
is an unsupervised machine learning problem used to identify sub-structure as communities
in such networks as computer and information infrastructures, social activities, and biological interactions or co-occurrences. It is used to find non-trivial topological features, with patterns of connection between nodes that are neither exactly regular nor random. For example, in metabolic networks, communities correspond to a series of chemical reactions called metabolic pathways~\cite{hanson2014metabolic}, whereas in a protein interaction network, communities correspond to proteins with similar functionality inside a biological cell~\cite{chen2006detecting}. In this work we focus on using Newman's modularity-based community detection~\cite{2006PNAS..103.8577N}.

QLS was applied to solving the
2-community detection problem on real networks of up to 410 nodes, while 
solving a 16 variable subproblem on a quantum device. To the best of our knowledge, this is the first attempt to tackle problems of this size using gate-model (universal) quantum computing. Also, QLS is shown to work with the D-Wave quantum annealer. We explore the potential of QLS as quantum devices become more and more capable and demonstrate its potential. 

The small size of available quantum devices creates a challenge, since typical algorithms (both quantum and classical) look at a problem ``as a whole", requiring large amounts of resources to store the description of the entire problem. While on classical computers storing the problem usually does not constitute a problem, it becomes a bottleneck when working with quantum computers that only have limited numbers of qubits and limited
connectivity between qubits. The number of variables that can be represented in a quantum
device is dependent on its underlying architecture.

A problem decomposition approach like local search presents a natural solution to this problem. A local search heuristic starts with some initial solution and searches its neighborhood iteratively, trying to find a better candidate solution with improved criterion (which is often an objective of the corresponding minimization or maximization of the problem). If a better solution is found, it replaces the current solution, and the search continues~\cite{aarts2003local}. Searching the neighborhood is a local problem and its size can be restricted to fit on a small quantum device. In QLS for graph community detection, the neighborhood of the solution is searched by selecting a subset of vertices and collectively moving them between the communities with the goal of improving the global modularity metric.

The QLS approach provides an additional benefit of being fundamentally hardware-agnostic. Local neighborhood search can be encapsulated as a routine, allowing researchers to easily switch between different hardware implementations. This is especially useful, since the landscape of quantum computing in the NISQ era is in a constant state of flux with many QC architectures available and new development happening constantly. It is not clear at this stage which architecture will become dominant in future. In this work we demonstrate how the two most developed and popular current paradigms, universal quantum computing (UQC) and quantum annealing (QA), can be integrated into the QLS framework and utilized to solve problems of practical size. Both paradigms have demonstrated great potential on a number of important  problems~\cite{king2018observation,romero2018strategies,ambainis2018quantum,dunjko2018computational}.

In this paper, we do not aim to analyze performance of quantum optimization algorithms like quantum annealing or QAOA. Although we do present some performance results (see Fig.~\ref{fig:gap}), they by no means constitute an exhaustive comparison with classical state-of-the-art. Instead, they provide motivation for our work, demonstrating that the subproblems offloaded to quantum solvers are not trivial and that hybridization is needed. For benchmarking, analysis and exploration the reader is referred to one of a number of recent paper analyzing QAOA performance~\cite{zhou2018quantum,guerreschi2018qaoa,osti_1492737}. In other words, we do not focus on finding and quantifying quantum speedups. Instead, we focus on a different question: if these algorithms are indeed capable of providing speedups in the near term, how can we leverage them to solve practical problems?

It is important to point out that the introduction of a problem decomposition heuristic like QLS limits the possible quantum speedup. Since to the best of our knowledge no asymptotic speedups have been shown so far for QAOA or QA, decomposition schemes limit the multiplicative speedup on the entire global problem by the multiplicative speedup on a small local subproblem. However, they still provide a way to take advantage of the small quantum devices that are becoming available. 

The rest of the paper is organized as follows. We begin by introducing the community detection problem and hybrid local-search schemes. Then we describe the QC paradigms we utilize and the quantum algorithms used to perform local search. Finally, we provide the implementation details, present the results and discuss their significance.

\section{The Community Detection Problem}

The community detection problem (or modularity network clustering) is an NP-hard problem~\cite{brandes2006maximizing} with a variety of applications in complex systems~\cite{newman2010networks}. 
Practical usefulness and complexity make community detection an interesting problem to tackle using QC. The goal of community detection in a network with an underlying simple undirected graph $G=(V,E)$ is to split the set of vertices $V$ into communities such that the modularity is maximized~\cite{2006PNAS..103.8577N}:

\begin{equation}
H = \frac{1}{4|E|}\sum_{ij}(A_{ij} - \frac{k_ik_j}{2|E|})s_is_j =  \frac{1}{4|E|}\sum_{ij}B_{ij}s_is_j,
\label{eq:mod}
\end{equation}

\noindent where the variables $s_i\in\{-1,+1\}$ indicate community assignment of vertex $i$ ($s_i=-1$ meaning vertex $i$ is assigned to the first community and $s_j=+1$ meaning that vertex $j$ is assigned to the second community), $k_i$ is a degree of $i\in V$, and $A$ is the adjacency matrix of $G$. In this work, we focus on clustering the network into two communities. There are several classical approaches to extend the problem to cases when the number of communities is greater than 2~\cite{2006PNAS..103.8577N,nascimento2011spectral}.

Community detection using a hybrid quantum-classical approach targeted for specific quantum architectures has been demonstrated previously. The 2-community problem was solved using \emph{qbsolv} and the D-Wave quantum annealer~\cite{ushijima2017graph} and extended for
$k$-communities~\cite{mniszewski2018,negre2019detecting}. Solving for 2-communities using QAOA and the IBM Q Experience was shown in~\cite{mniszewski2018}. Solving for $k$-communities on signed graphs using block coordinate descent~\cite{bcd1, bcd2} and D-Wave quantum annealer was shown in~\cite{zahedinejad2019multi}.

\section{Quantum-accelerated Decomposition Heuristics for Optimization}

Central to the discipline of QC in the NISQ era is the problem of a limited number of available noisy qubits. For example, at the time of writing, the largest gate-model QC device available on the cloud was IBM Q 20 Tokyo~\cite{ibmqxdevices} with twenty superconducting qubits. Twenty qubits translates into up to 20 variables due to connectivity constraints. This implies that the maximum number of nodes of a network we can cluster directly is 20. This example highlights the challenges of leveraging limited NISQ-era devices to solve practical problems and motivates our local-search approach. Note that same considerations apply for problems other than optimization. For example, similar hybrid approaches have been applied to Blind Quantum Computation~\cite{giovannetti2013efficient, barz2012demonstration, sheng2015deterministic, sheng2018blind}, and distributed quantum machine learning~\cite{sheng2017distributed}. Parallel Quantum Computation (PQC)~\cite{long2004parallel} can be used to speed up Grover's search algorithm~\cite{grover1996fast} by dividing a database on which the search is performed between an ensemble of quantum computers running in parallel~\cite{xiao2002fetching,long2003experimental}.  

In response to the challenges of quantum computation in the NISQ era, a number of decomposition approaches have been explored. The methods described in this section use limited in size quantum optimization solvers to search a restricted neighborhood of a given solution with the goal of finding a better solution. Here the given solution comes either from running a classical heuristic solver on a CPU or from the previous iteration.
These methods are inspired by the success of classical large-scale  neighborhood local search methods (the reader is referred to~\cite{rotta2011multilevel} for a survey of local-search heuristics in general and to~\cite{ahuja2002survey} for a survey of large-scale neighborhood methods in particular). It is important to note that unlike this paper, all the works described in this section focus exclusively on D-Wave quantum annealers. 

The first family of methods builds on classical pre-processing methods for quadratic unconstrained binary optimization (QUBO) problems (see~\cite{tavares2008new} for a review). One such pre-processing technique is heuristically fixing variables. The variables are chosen by maintaining a set of elite solutions and fixing the variables that have the same value across many or all local optima, with the intuition being that they will have the same values for the global optimum~\cite{wang2011effective}. Sample persistence variable reduction (SPVAR)~\cite{karimi2017boosting} in its basic version uses a sample of solutions (obtained either from a quantum annealer or a classical heuristic) and fixes the variables that have the same value across the entire sample. Then SPVAR uses a quantum annealer as the solver for the restricted QUBO. This method was later extended by introducing multistart (multiple samples) and was extensively benchmarked using both the D-Wave quantum annealer as well as state-of-the-art classical heuristics for Chimera Hamiltonians~\cite{karimi2017effective}.

The second family of methods extends iterative large-scale neighborhood local search methods. Local search commonly considers the neighborhood of bit strings that have Hamming distance one from the current solution at each step. The performance of local search methods can be improved by considering larger neighborhoods (Liu et al.~\cite{liu2005hybrid} shows significant performance improvements for neighborhood of Hamming distance four, equivalent to fixing all but four variables). Quantum optimizers provide a potentially efficient way to explore these larger neighborhoods. This rather straightforward idea was introduced in~\cite{bian2014discrete} and extended and rigorously tested in~\cite{bian2016mapping, rosenberg2016building, zintchenko2015local}. A similar hybrid tree search method was presented in~\cite{tran2016hybrid}. These methods utilize the D-Wave quantum annealer as the quantum optimizer, enabling them to solve problems with thousands of variables. In this work we limit the subproblem size to be small enough to fit on the IBM Q quantum computer, limiting the size of the problems we can tackle.
D-Wave provides a set of utilities for problem decomposition, including a hybrid extension of the tabu search QSage~\cite{dwave-handbook}. 

\section{Quantum Local Search}

To address the challenges outlined above, we introduce the QLS algorithm. QLS is a hybrid quantum-classical local-search approach, inspired by numerous existing local-search heuristics. QLS is motivated by the successful application of local-search heuristics to a variety of optimization problems. 
The novelty of QLS is that it can utilize both quantum annealers and universal quantum computers.
In this work, we apply QLS to the problem of 2-community detection on graphs, but the success and versatility of local-search heuristics make us confident that QLS can be extended to other optimization problems.

In QLS for community detection, the local search starts with a random assignment of communities to vertices
and attempts to iteratively optimize the current community assignment of a \emph{subset} of vertices with the goal of increasing modularity. Here the space of potential community assignments of a subset of vertices plays the role of the neighborhood where the local search is performed. At each iteration, a subset $X\subset V$ is populated by selecting vertices with the highest potential gain in modularity obtained when changing their community assignment. This can be done efficiently~\cite{2006PNAS..103.8577N} since at each iteration we only need to update the gains of vertices in $X$ and their neighbors. Then at each iteration, the community assignment of the vertices in the subset $X$ (subproblem) 
is optimized using a routine that includes a call to a quantum device.
The local search proceeds until it converges. We define convergence as three iterations with no improvement in modularity. Note that in general it is not necessary to consider all vertices before convergence: in the 2-community problem, random initial assignment would be correct for 50\% of vertices on average. Our approach is outlined in Algorithm~\ref{alg:outline}.

\begin{algorithm}[H]
 \caption{QLS Community Detection}\label{alg:outline}
\begin{algorithmic}
\State solution = initial\_guess($G$)
 \While{not converged}
  	\State $X$ = populate\_subset($G$)
  	\State // \textit{using IBM UQC or D-Wave QA}
  	\State candidate = solve\_subproblem($G$, $X$)
  	\If{$\mbox{candidate} > \mbox{solution}$}
  	\State solution = candidate
    \EndIf
 \EndWhile
\end{algorithmic}
\end{algorithm}

The subproblem of optimizing community assignment of the subset is formulated by fixing community assignment for all vertices not in the subset ($i\not\in X$) and encoding them into the optimization problem as boundary conditions. This is a commonly used technique in many heuristics~\cite{leyffer2013fast,hager2018multilevel}. 
Denoting fixed assignments by $\tilde{s}_j$, the subproblem can be formulated as:

\begin{equation}
\label{eq:subproblem}
\arraycolsep=1.4pt
\begin{array}{r c l}
Q_{s} & = & \sum_{i>j | i,j\in X}2B_{ij}s_is_j +  \sum_{i\in X}\sum_{j\not\in X}2B_{ij}s_i\tilde{s}_j \\
 & = & \sum_{i>j| i,j\in X}2B_{ij}s_is_j + \sum_{i\in X}C_{i}s_i, \\
 \mbox{where } C_{i} & = & \sum_{j\not\in X}2B_{ij}\tilde{s}_j
\end{array}
\end{equation}

Clearly, maximizing~(\ref{eq:subproblem}) can only increase global modularity~(\ref{eq:mod}). The objective defined in Eq.~(\ref{eq:subproblem}) can be optimized using a QC algorithm. The exact way the optimization is performed can vary between different QC implementations, making our approach extendable to new emerging QC platforms. We demonstrate this portability by implementing two subproblem optimizing routines that use IBM Q 16 Rueschlikon~\cite{ibmqxdevices} and D-Wave 2000Q~\cite{dwave2018}. Additionally, we implement a subset optimization routine that uses the classical Gurobi solver~\cite{optimization2014inc} for quality comparison. The choice of Gurobi is not of importance, since for subproblems with 16 variables any classical integer programming solver is capable of finding the optimum.

\section{Quantum Computing Paradigms}
Quantum annealing (QA) is a form of adiabatic quantum computation (AQC)~\cite{mcgeoch2014adiabatic}. 
QA solves an optimization problem by encoding it as an Ising model Hamiltonian, with the ground state of that Hamiltonian corresponding to the global solution
of the optimization problem. The Ising Hamiltonian describes the energy of a collection of $n$ spin variables, with each variable being in one of two spin states ($\pm1$). A spin configuration describes assignment of states to spin variables, with $s_i$ denoting the state of
spin variable $i$ (note that the 2-community problem maps naturally to this system, with the resulting spin state, $s_i$, denoting community assignment). The energy of a configuration is then defined by:
\begin{equation}
    H(s) = \sum_{i>j}J_{ij}s_is_j+\sum_{i}h_is_i
\end{equation}
\noindent where $h_i$ correspond to external forces applied to spin variables, and $J_{ij}$ to coupling strengths between pairwise spin interactions~\cite{mcgeoch2014adiabatic}.

QA finds the ground state of the objective Hamiltonian by performing a quantum evolution. As the initial Hamiltonian, QA uses a transverse field Hamiltonian. It introduces quantum fluctuations that help the annealing process to escape local minima by ``tunneling through" hills in the energy landscape, enabling the evolution to move faster than adiabatic requirement would allow. As the evolution is performed, the transverse field Hamiltonian is slowly ``turned off" (scaled with a coefficient decreasing to 0), such that the evolution finishes in a system described by the problem Hamiltonian~\cite{mcgeoch2014adiabatic}.

Since AQC was introduced in 2000 by Farhi et al.~\cite{farhi2000quantum}, D-Wave Systems Inc~\cite{dwave2018}, IARPA's QEO effort~\cite{iarpaqeo} and other researchers~\cite{novikov2018exploring}
have achieved a lot of progress in developing a system implementing QA~\cite{mcgeoch2014adiabatic} and applying it to a variety of problems, including optimization problems on graphs \cite{ushijima2017graph}, machine learning \cite{omalley2017}, traffic flow optimization \cite{neukart2017}, integer factoring~\cite{peng2019factoring} and simulation problems \cite{harris2018}.
Optimization problems can be solved by QA when formulated in the Ising form
(\ref{eq:subproblem}) or as a quadratic binary optimization (QUBO).

Universal (or gate-based) quantum computing was introduced in the 1980s~\cite{deutsch1985quantum} and has seen great theoretical advances since. Shor's~\cite{shor1994algorithms} and Grover's~\cite{grover1996fast} algorithms are two most famous examples of quantum algorithms with theoretically proven speedups over classical state-of-the-art. Universal quantum computing has been implemented in hardware by a number of companies, national laboratories and universities~\cite{ballance2016high,barends2014superconducting, petta2005coherent, acin2018quantum,saffman2010quantum}.

To optimize (\ref{eq:subproblem}) on a universal quantum computer, we use a hybrid
quantum-classical approach, Quantum Approximate Optimization Algorithm (QAOA)~\cite{farhi2014quantum,farhi2014quantumbounded}. Similar to QA, a problem is encoded as an objective Hamiltonian $H$. Then a quantum evolution is performed starting with some fixed initial easy-to-prepare
state (traditionally, uniform superposition over computational basis states is used). The difference is that unlike QA, in which the evolution is analog, in QAOA the evolution is performed by applying a series of gates parameterized by a vector of variational parameters $\theta$. 
A hybrid approach, combining the quantum device performing the evolution and a classical optimizer, finds the optimal variational parameters. QAOA starts with an initial set of variational parameters $\theta_0$. At each step, a multi-qubit state $\ket{\psi{(\theta)}}$ parameterized by the variational parameters $\theta$ is prepared on the quantum co-processor. Then a cost function $E(\theta) = \bra{\psi{(\theta)}}H\ket{\psi{(\theta)}}$ is measured and the result is used by the classical optimizer to choose new parameters $\theta$ with the goal of finding the ground-state energy $E_G=\min_{\theta}\bra{\psi{(\theta)}}H\ket{\psi{(\theta)}}$. QAOA provides a viable path to quantum advantage~\cite{farhi2016quantum}, making it a good algorithm to explore on near-term quantum computers.

\section{Results and discussion}
We implement the classical part of QLS in Python 3.6, using NetworkX~\cite{hagberg2008} for network operations. 
The subproblem solvers are implemented using QA (D-Wave SAPI), QAOA (IBM QISKit~\cite{cross2018ibm}) and the classical Gurobi solver~\cite{optimization2014inc}. Our framework is modular and easily extendable, allowing researchers to add new subproblem solvers as they become available. The framework is available on GitHub at \url{http://bit.ly/QLSCommunity}.

\begin{figure}
\centering
    \includegraphics[width=1\textwidth]{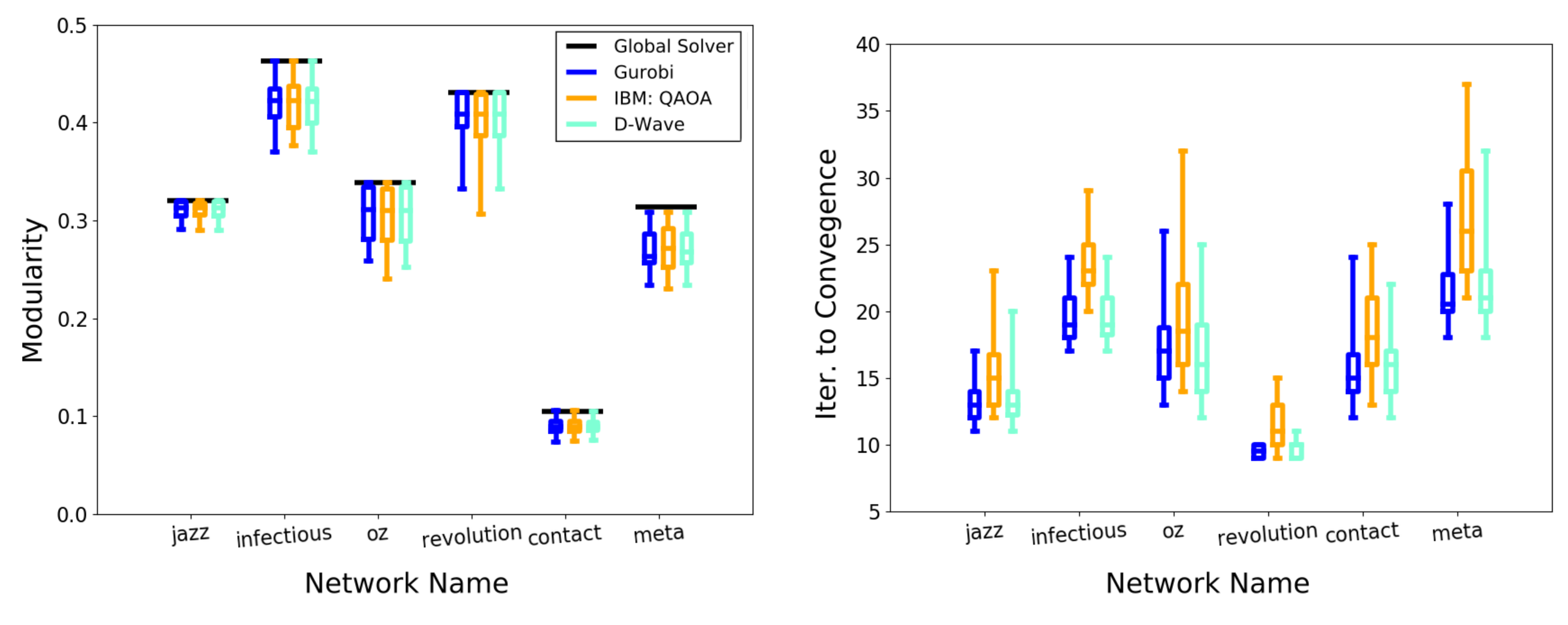}
    \caption{Box-plots showing the range of modularity scores for 2-community detection (left, greater is better) and number of solver calls (right, less is better) respectively for the three different subproblem solvers. The results show that the proposed approach is capable of achieving results close to the state-or-the-art (Global Solver)}
    \label{fig:results}
\end{figure}

\begin{figure*}[ht]
    \centering
    \includegraphics[width=\textwidth]{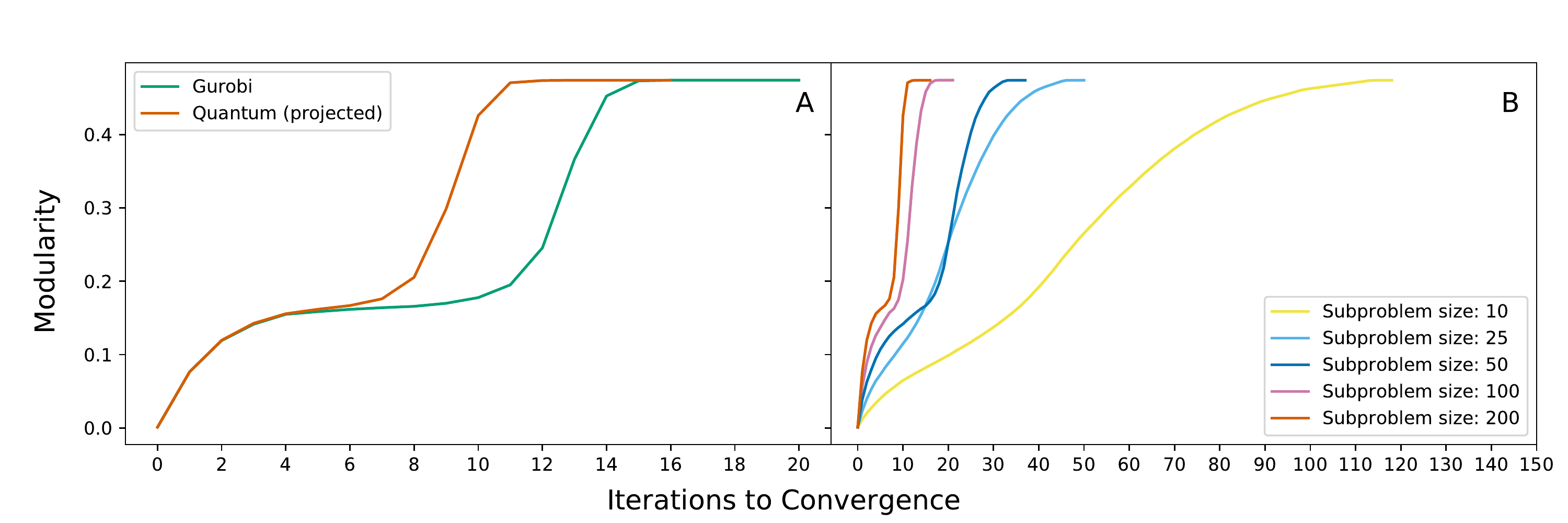}
    \vspace{-0.35cm}
    \caption{A projection of QLS performance as the hardware size of quantum devices increases. \textbf{(A)} Projected of QLS performance as the quality of the local search solver solution is improved. The projection is performed by comparing the performance of classical solver Gurobi with time limit fixed at 0.25s (D-Wave time to solution) and Gurobi with time limit 1000s (projected good solution). The assumption is that the new quantum optimization algorithms would be able to scale and provide results of the same quality as Gurobi with time limit 1000s while taking approximately the same time to solve the problem as they do today. \textbf{ (B)} Projected number of iterations for QLS to converge as larger devices become available (projection performed by using Gurobi as a subproblem solver).}
    \label{fig:gap}
    \vspace{-0.25cm}
\end{figure*}

In order for a subproblem to be solved on the D-Wave system, the problem is embedded onto the physical layout (Chimera graph). The clique embedder \cite{boothby2016fast} is used to calculate an embedding of a complete 16-variable problem once and is reused for each subproblem. In this work, we utilized D-Wave's Solver API (SAPI) which is implemented in Python 2.7, to interact with the system. We used the D-Wave 2000Q which has up to 2048 available qubits. Subproblems of approximately 64 variables can be solved on the the 2000Q, however, for a fair comparison, we limit ourselves to up to 16 variables. The D-Wave system is intrinsically a stochastic system, where solutions are sampled from a distribution corresponding to the lowest energy state. For each subproblem, the best solution out of 10,000 samples is returned. 

The QAOA subproblem solver is implemented using the IBM QISKit framework. We ran QAOA with RYRZ ansatz~\cite{ibmqryrz} on the IBM 16 Q Rueschlikon~\cite{ibmqxdevices} with 16 qubits. For optimization of the variational parameters we used a SciPy~\cite{scipy} implementation of Constrained Optimization BY Linear Approximation (COBYLA) method~\cite{powell1994direct}. For each subproblem, we performed optimization of the variational parameters $\theta$ using a high-performance simulator~\cite{ibmqsimulator} and ran QAOA with optimized parameters on a quantum device using the IBM Q Experience~\cite{ibmqxdevices} cloud service. We allowed COBYLA 100 function evaluations (i.e. 100 QAOA runs on the simulator) to find optimal parameters $\theta$. We used this setup (training on a simulator and running on the quantum device) because of the limitations of the IBM Q Experience job queue at this time. In our experience, jobs submitted to the IBM quantum device can spend minutes to hours in queue, requiring days to complete a full variational parameter optimization loop. It is our understanding that this will be remedied in the future. The main downside of this setup is that the variational parameters trained on a simulator do not encode the noise profile of the device, decreasing the quality of the solution. This is one of the main factors contributing to slightly slower convergence for QAOA compared to other methods. In the future, as various QC devices become available, it will be straightforward to perform QAOA fully on a QC device. However, even using the current setup we achieved very promising results, indicating great potential for applying variational quantum-classical methods to combinatorial optimization problems.

Our results are presented in Fig.~\ref{fig:results}. We ran our algorithm on six real-world networks from the KONECT dataset~\cite{kunegis2013konect} with up to 410 nodes as our benchmark. The networks come from different real world phenomena and include social and metabolic networks. For each network, we ran 30 experiments with different random seeds. The same set of seeds was used by the three subproblem solvers, with all solvers starting with the same initial guess and therefore making the results directly comparable. We fixed the subproblem size at 16 vertices. Our results demonstrate that QLS with both D-Wave QA and QAOA on IBM Q as quantum subproblem solvers perform similarly in terms of quality of the solution (modularity) and the number of iterations to convergence, and are capable of achieving results comparable to state-of-the-art. Our results are compared to results using the Gurobi Optimizer, which is a state-of-the-art solver for mathematical programming. We use the Gurobi Optimizer in two ways: first as a solver for solving the entire problem at once, which we report as the \emph{Global Solver} and second as a solver for solving small size subproblems of fixed size within the local search framework. For solving the entire problem, the Gurobi Optimizer is unable to reach a provable global optima for most of the problems within the specified time-frame. For the graph problems of up to approximately 400 variables, we run Gurobi (as a global solver) for up to 72 hours and the results reported are within an optimality gap of up to 33\%. For the smaller size subproblems of 16 variables, Gurobi was able to find the optimal solution within less than second. The networks and the set of seeds we used are available online at \url{http://bit.ly/QLSdata}.

The results demonstrate the promise of the proposed approach. We presented a framework that is able to find 2 communities in graphs of size up to 410 vertices using only NISQ-era devices. We explored the potential of our approach as new and better QC hardware becomes available in two ways. First, we used the classical Gurobi solver~\cite{optimization2014inc} to simulate the performance improvements in QLS as the subproblem size is increased (see Fig.~\ref{fig:gap}B). We generate a 2000 node random graph with realistic community structure and known modularity~\cite{sah2014exploring}.  Unsurprisingly, QLS finds the optimal solution faster (using fewer local search iterations) as the subproblem size increases. Second, we demonstrate the need for quantum acceleration by demonstrating the limitations of existing state-of-the-art solvers. We used Gurobi~\cite{optimization2014inc} as a subproblem solver with subproblem size of 200. Fig.~\ref{fig:gap}A shows that for the subproblem of this size, Gurobi cannot produce a good solution quickly. We compared Gurobi with time limit 0.25s (the running time of QA on D-Wave) with Gurobi with time limit 1000s, with the assumption that Gurobi would converge to a good solution. We use the running time of QA as our estimate because at the time of writing we do not have a good way of measuring the running time of QAOA due to the architecture of the IBM Q Experience. We expect QAOA to have similar performance. This assumes that quantum methods would scale well to larger problems, which is a strong assumption. However, the goal here is to motivate the exploration of quantum optimization heuristics by showing the limitations of classical state-of-the-art and not to demonstrate quantum advantage. Using a better solution within the local search enables 25\% (4 iterations) improvement in time to convergence (convergence is defined as three iterations with no improvement). This demonstrates that the subproblems become computationally hard even for sizes that are small enough to potentially fit on near-term devices. It is important to note that even though in our experiments Gurobi performed better than other integer programming solvers, it is quite possible that other solvers can perform better on this problem, especially after tuning. Indeed, in the past the improvements in classical heuristics have forced researchers to downgrade claims of quantum advantage~\cite{dwavenoadv1, dwavenoadv2}. However, demonstrating quantum advantage is outside of scope of this paper. Instead, we use these results to motivate our hybrid approach by showing the computational complexity of the subproblems offloaded to quantum solvers. As quantum solvers improve and become capable of providing speedups at \emph{subproblem} level, out QLS will enable us to leverage these speedups at the global problem level. 

\section{Conclusion}
In the next few years a number of QC hardware implementations are expected to become mature enough to be applied to practically important problems. QC using trapped ions~\cite{chmielewski2018cloud} and Rydberg atom arrays~\cite{pichler2018quantum} are just two examples of quantum hardware now moving out of the laboratory, with the potential to realize quantum advantage. However, none of them promise to deliver more than a few hundred qubits in the near future. Therefore, we believe the future of QC is hybrid, with algorithms combining both classical and quantum computation. QLS presents a path to integrate NISQ-era devices into computational workflows in a flexible way, both in terms of adding different hardware backends and extending to different problems. Classical local search heuristics have proven useful for a variety of problems in many fields~\cite{rotta2011multilevel}. We believe that QLS can be similarly extended to problems beyond network community detection. We also believe that the decomposition approaches like QLS can improve dramatically the speed and performance of QAOA algorithms on universal quantum computers, which might the key to achieve quantum advantage on NISQ devices.

\smallskip
\section{Acknowledgements}
This research used quantum computing system resources of the Oak Ridge Leadership Computing Facility, which is a DOE Office of Science User Facility supported under Contract DE-AC05-00OR22725. This research used the resources of the Argonne Leadership Computing Facility, which is a U.S. Department of Energy (DOE) Office of Science User Facility supported under Contract DE-AC02-06CH11357. Yuri Alexeev and Ruslan Shaydulin were supported by the DOE Office of Science. We gratefully acknowledge the computing resources provided and operated by the Joint Laboratory for System Evaluation (JLSE) at Argonne National Laboratory.  The authors would also like to acknowledge the NNSA’s Advanced Simulation and Computing (ASC) program at Los Alamos National Laboratory (LANL) for use of their Ising D-Wave 2000Q quantum computing resource. LANL is operated by Triad National Security, LLC, for the National Nuclear Security Administration of U.S. Department of Energy (Contract No. 89233218NCA000001). Clemson University is acknowledged for generous allotment of compute time on Palmetto cluster.

\bigskip

\bibliography{qaoa,ilya}

\end{document}